\documentclass[conference]{IEEEtran}
\IEEEoverridecommandlockouts
\usepackage{cite}
\usepackage{amsmath,amssymb,amsfonts}
\usepackage{algorithmic}
\usepackage{graphicx}
\usepackage{textcomp}
\usepackage{xcolor}
\usepackage{cases}
\usepackage{makecell}
\usepackage{hyperref}
\usepackage{multirow}
\usepackage{flushend}
\def\BibTeX{{\rm B\kern-.05em{\sc i\kern-.025em b}\kern-.08em
    T\kern-.1667em\lower.7ex\hbox{E}\kern-.125emX}}
\begin{document}

\title{
\vspace{0pt}
\begin{center}
\Huge
Scorpiano -- A System for Automatic Music Transcription for Monophonic Piano Music
\end{center}
}

\author{\IEEEauthorblockN{Bojan Sofronievski and Branislav Gerazov}
\IEEEauthorblockA{
    \textit{Faculty of Electrical Engineering and Information Technologies} \\
    \textit{Ss Cyril and Methodius University in Skopje, Macedonia} \\
    \textit{bojan.sof@hotmail.com, gerazov@feit.ukim.edu.mk}
}}

\maketitle
\begin{abstract}
Music transcription is the process of transcribing music audio into music notation.
It is a field in which the machines still cannot beat human performance.
The main motivation for automatic music transcription is to make it possible for anyone playing a musical instrument, to be able to generate the music notes for a piece of music quickly and accurately.
It does not matter if the person is a beginner and simply struggles to find the music score by searching, or an expert who heard a live jazz improvisation and would like to reproduce it without losing time doing manual transcription.
We propose Scorpiano -- a system that can automatically generate a music score for simple monophonic piano melody tracks using digital signal processing. 
The system integrates multiple digital audio processing methods: notes onset detection, tempo estimation, beat detection, pitch detection and finally generation of the music score.
The system has proven to give good results for simple piano melodies, comparable to commercially available neural network based systems.
\end{abstract}

\begin{IEEEkeywords}
automatic music transcription, musical note recognition, pitch detection, onset detection, audio processing
\vspace{10mm}
\end{IEEEkeywords}

\section{Introduction}

Automatic music transcription (AMT) is a process that automatically identifies the performed notes in a given melody track, with the goal of generating a music score.
Besides automatic generation of music scores, AMT applications include interactive music systems and automated music tutors that teach playing an instrument \cite{AMT:1, AMT:2}.

One of the main determinants for the difficulty of the task of AMT is whether the music to be transcribed is monophonic or polyphonic.
Monophonic music simply means that the performer plays only one note at a time. 
There must not be sounds which overlap and this sounds are characterized by only one pitch.
Contrary to monophonic music, polyphonic music consists of two or more simultaneous lines of independent melodies.
This means that multiple notes are played at the same time and this sound is characterized with multiple pitches.
AMT for polyphonic music is a non-trivial task and is a very active field of research \cite{AMT:2, AMTpoly:1, AMTpoly:2}.

There are few different approaches to AMT.
One approach is based purely on digital signal processing and it is mainly used for monophonic music. 
These systems basically integrate methods for pitch and onset/beat detection.
There are mainly two approaches for pitch detection: \emph{i}) the time-domain approach, which is primarily based on the autocorrelation function and its modifications \cite{PITCHm:1, YIN:1}, \emph{ii}) the frequency-domain approach, which is primarily based on the STFT (Short Time Fourier Transform) \cite{PITCHf:1}.
The onset detection algorithms are based on finding the peaks of a novelty function, i.e., a function whose peaks should coincide, within a tolerance margin, with note onset times \cite{ONSET:2}.

The second approach is based on machine learning algorithms, mainly neural networks and these systems are the main choice for transcribing polyphonic music \cite{AMTML:1}.
There are a few commercially available applications which follow this approach, such as AnthemScore\footnote{\url{https://www.lunaverus.com}} and Melody Scanner\footnote{\url{https://melodyscanner.com}}, and they advertise over 80\% accuracy.
However, neural networks in general require more processing power and a big set of data for training.

We propose the system for AMT of monophonic music, based on the digital signal processing approach, named Scorpiano.
The piano is chosen to be the source instrument for transcription, because it produces sound by hitting the strings with hammers, giving the piano better frequency stability of the pitch, compared to other string instruments which produce sounds by plucking the strings.
Our systems generates scores with high accuracy and fast transcription speeds, and is comparable to commercially available neural network based systems.
The system also only has a small number of parameters that can be easily adjusted for different piano sources.
Scorpiano is made available as free software.\footnote{\url{https://gitlab.com/BojanSof/scorpiano}}

\section{System Architecture}

\begin{figure}[t]
\centering
\vspace{-10pt}
\includegraphics[clip, width=3.5in]{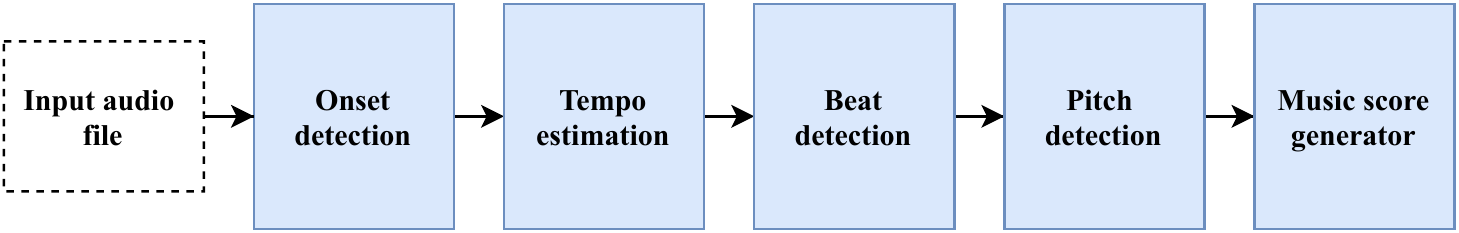}
\caption{System architecture.}
\label{fig_dgm}
\vspace{-2pt}
\end{figure}

Fig.~\ref{fig_dgm} shows Scorpiano's system architecture. 
The input to the system is an audio file of a monophonic piano recording.
The system consists of five modules for: onset detection, tempo estimation, beat detection, pitch detection and music score generation.
The system outputs an image file of the generated music score.

\subsection{Onset Detection}

\begin{figure}[bt]
\centering
\vspace{-10pt}
\includegraphics[clip, width=3.5in]{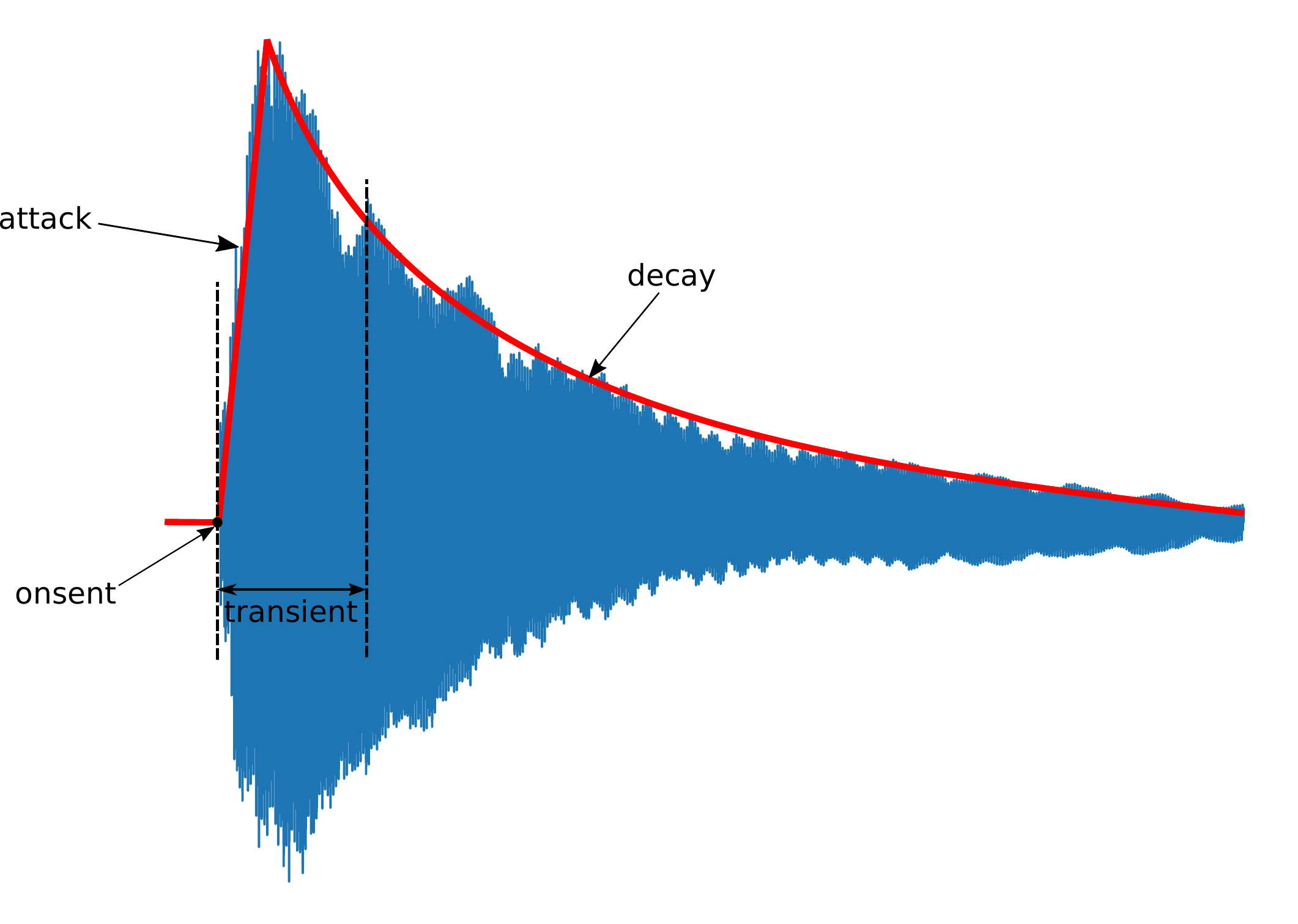}
\caption{The onset, attack, transient, and decay of an isolated note.}
\label{fig_envelope}
\vspace{-2pt}
\end{figure}

Onset detection is the task of determining the starting times of musical notes in a music recording.
Onsets correspond to a sudden increase of energy at the beginning of musical notes.
There are a number of closely related concepts defined for each note realisation~\cite{ONSET:1}, as illustrated in Fig.~\ref{fig_envelope}:
\begin{itemize}
    \item the \emph{onset} of the note is the time moment when the transient starts,
\item the \emph{attack} of the note is the time interval during which the amplitude envelope increases, and
\item the \emph{transient}, in the case of the piano or other acoustic instruments, corresponds to the period during which the excitation is applied and then damped, leaving only the slow decay at the resonance frequency of the body of the instrument, and
\item the \emph{decay} is defined as the time interval in which the amplitude decreases gradually until the sound vanishes.
\end{itemize}

To detect a sudden increase of the energy, we compute the energy novelty function. 
The energy novelty function is a function that describes local changes in signal energy. 
To compute the novelty function, first we compute the local energy function of the signal $x$, using a bell-shaped window function $w$ with length $2M+1$, e.g. a Hann window, given with:
\setlength{\arraycolsep}{0.0em}
\begin{eqnarray*}
E[n] &{}={}& \sum \limits_{m=-M}^{m=M} |x[n+m] w[m]|^2 \nonumber \\
     &{}={}& \sum \limits_{m \in Z} |x[m]w[n-m]|^2  \nonumber \\
     &{}={}& x^2[n] * w^2[n] \nonumber
\end{eqnarray*}
\setlength{\arraycolsep}{5pt}
\\

To compute the changes of the energy, we take the first derivative of the local energy, which in the discrete case can be approximated by taking the difference between subsequent energy values.
Before taking the derivative, we apply logarithmic compression to the local energy function, taking into account the fact that human perception of sound intensity is logarithmic.
Because we are interested only in energy increases, we apply a half-wave rectification of the derivative.
The half-wave rectification function is given by:
\setlength{\arraycolsep}{0.0em}
\begin{numcases} {r_{\rm{half}}(x)=}
x, & \rm{if} $x \geq 0$ \nonumber \\
0, & \rm{if} $x < 0$ \nonumber
\end{numcases} 
\setlength{\arraycolsep}{5pt}

Thus, we obtain the local energy novelty function:
\begin{displaymath}
E_{\rm{novelty}}[n]=r_{\rm{half}} \left(E_{\gamma}[n+1]-E_{\gamma}[n] \right) 
\end{displaymath}
where $E_{\gamma}[n]$ is the logarithmic compression of $E[n]$ for a positive constant $\gamma$:
\begin{displaymath}
E_{\gamma}[n] = \log \left( 1 + \gamma \cdot E[n] \right) 
\end{displaymath}

\begin{figure}[bt]
\centering
\vspace{-10pt}
\includegraphics[clip, width=3.5in]{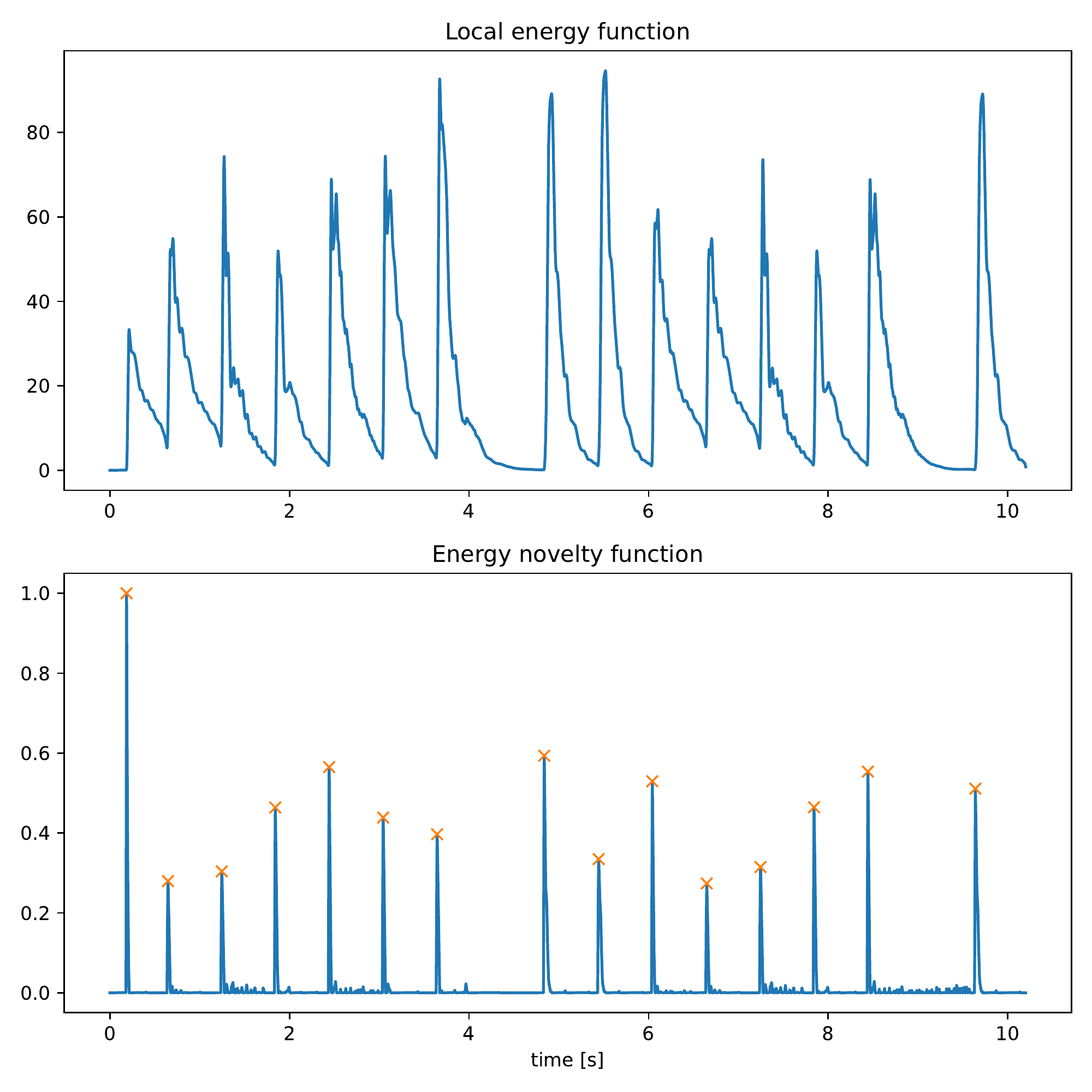}
\caption{Local energy function and energy novelty function.}
\label{fig_onset}
\vspace{-2pt}
\end{figure}

Fig.~\ref{fig_onset} shows the local energy, the energy novelty function and the marked maxima of the novelty function for a part of ``Twinkle twinkle little star'' melody, using a Hann window with a window length of 46 ms.
Note that the energy novelty function is normalised by dividing it with its maximum value.

To detect onsets, we mark the local maxima of the energy novelty function.
A simple method for finding local maxima is employed, keeping only maxima above a set amplitude threshold, and discarding maxima that are two close together. 
We set the amplitude threshold to be 0.1, and the time threshold to be 0.1~s.

\subsection{Tempo Estimation}

To find the correct timing of the musical notes and their beat duration, we must estimate the tempo of the music.
For this purpose, we use the \texttt{beat.tempo} function from \emph{librosa}\footnote{\url{https://librosa.org}} library \cite{LIBROSA} to obtain the beats per minute value for the melody. 

\subsection{Beat Detection}

After detecting the onsets of the musical notes and estimating the tempo, the beat duration of each note should be calculated.
For every note, except the last, assuming there are no breaks, note duration can be computed by taking the difference between two subsequent onset moments.
Then, the beat duration of the note is computed by multiplying the note duration with the tempo.
To find the duration of the last note we can simply track backwards the normalised energy of the signal which represents the last note, and estimate the end of the note using a threshold.

\subsection{Pitch Detection}

According to the Fourier representation, each musical sound is a weighted sum of an infinite number of sinusoidal components.
The frequency values of this components are integer multiples of the first one, called the fundamental frequency, denoted as $F_0$, which is perceived as pitch.

We can apply pitch detection using the notes' starting and ending times determined by their onset.
We use the YIN algorithm for pitch detection~\cite{YIN:1}.
It is an improved autocorrelation method for pitch detection, with a few modifications to minimize the error.
The YIN algorithm can be sublimed in 6 steps:
\begin{enumerate}
    \item calculate the autocorrelation function:
\begin{displaymath}
r_t [\tau] = \sum \limits_{j=t+1}^{t+M-\tau} x[j] x[j+\tau]
\end{displaymath}
    \item calculate the difference function, over a window with size M samples (corresponding to 68 ms window length in our case):
\begin{displaymath}
d_t [\tau] = \sum \limits_{j=1}^{M} \left( x[j] - x[j+\tau] \right)^2
\end{displaymath}
which can be expressed using the autocorrelation function as:
\begin{displaymath}
d_t [\tau] = r_t[0]+r_{t+\tau}[0]-2r_t[\tau]
\end{displaymath}
    \item calculate the cumulative mean normalized difference function:
\setlength{\arraycolsep}{0.0em}
\begin{numcases} {d'_t[\tau]=}
1, &\rm{if} $\tau=0$ \nonumber \\
\frac{d_t[\tau]}{\left( 1/\tau \right) \sum\limits_{j=1}^{\tau}d_t[j]}, & otherwise \nonumber
\end{numcases}
\setlength{\arraycolsep}{5pt}
    \item threshold -- set the absolute threshold and choose the smallest value of $\tau$ that gives minimum of the cumulative mean normalized difference function, smaller than the threshold,
    \item calculate parabolic interpolation -- each local minimum of the cumulative mean normalized difference function and its neighbors is fit by a parabola and the abscissa of the interpolated minimum is used as the period estimate.
    \item calculate the best local estimate -- repeat the period estimation in a shrinking time interval.
\end{enumerate}

To estimate the fundamental frequency, the median of the estimated pitch periods for each window is computed, due to the fact that the median is more immune to single or few erroneous extreme values in the list of numbers, compared to the average of the numbers.

\begin{figure}[bt]
\centering
\vspace{-10pt}
\includegraphics[clip, width=3.5in]{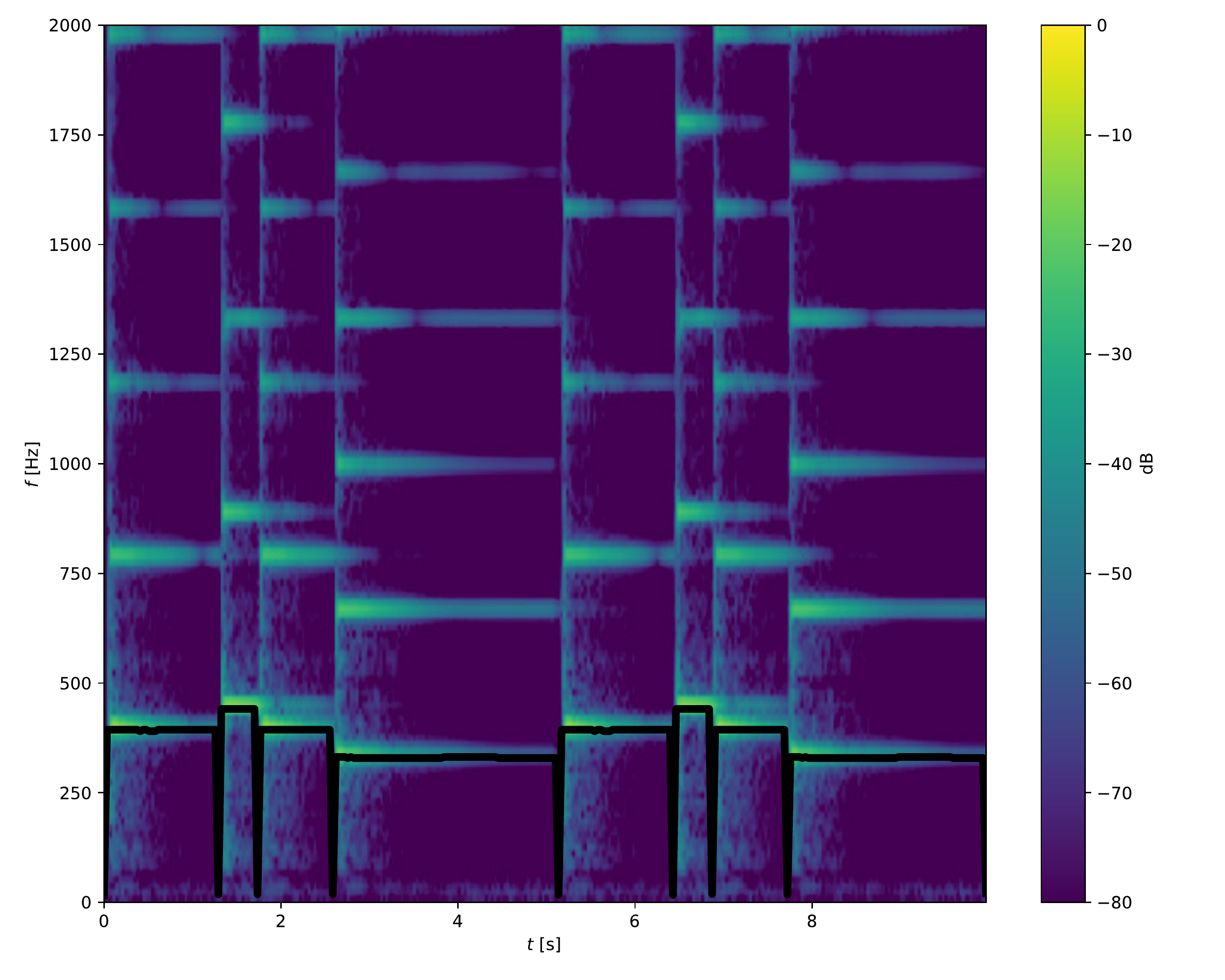}
\caption{Pitch contour obtained with the YIN algorithm for part of ``Silent Night``}
\label{fig_pitch}
\vspace{+2pt}
\end{figure}

Figure \ref{fig_pitch} shows the pitch contour obtained with the YIN algorithm for part of the ``Silent Night'' melody.
The peaks at beginning and ending of each note show why the fundamental frequency is computed using the median of the estimated pitch periods for each window.

\subsection{Music Score Generator}

Using the obtained beat durations and pitch of the musical notes, we can generate the music score.
The library used for this step is \emph{music21}\footnote{\url{https://web.mit.edu/music21}}, with the \emph{LilyPond} engine\footnote{\url{http://lilypond.org}}.
The time signature of the score must be given by the user.
The output is an image file.

\section{Experiments}

To evaluate the performance of the system, ten popular monophonic piano melodies are chosen.

In the first experiment, the system is tested with the melodies generated using the \emph{MuseScore}\footnote{\url{https://musescore.org}} application, which allows the generation of audio files from the music score for the melody.

In the second experiment, the system is tested with recordings of melodies played on a Miditech Midistart 3 MIDI keyboard, capable of modelling the dynamics of a piano.
For simulating the piano sound, the \emph{Addictive Keys}\footnote{\url{https://www.xlnaudio.com/products/addictive_keys}} software was used and the audio output of the program was recorded using \emph{Audacity}\footnote{\url{https://www.audacityteam.org}}.
We will refer to this configuration as ``digital piano''.

In the third experiment, the effect of the change of tempo on the system accuracy is evaluated with two melodies played on the digital piano, with three different tempos: slow, normal and fast, i.e. 80 bpm, 100 bpm and 120 bpm.

The performance of our system is compared with the commercial \emph{AnthemScore} program.
AnthemScore is an AMT software based on neural networks, advertised as being trained on millions of data samples.

To evaluate the performance of our system we define the following three errors:
\begin{itemize}
	\item \textbf{note error rate} -- the relative error for the number of original, $n_{\rm{original}}$, and detected notes, $n_{\rm{detected}}$:
\begin{displaymath}
\varepsilon_{\rm{note}} = \frac{|n_{\rm{detected}} - n_{\rm{original}}|}{n_{\rm{original}}} \times 100 \%
\end{displaymath}
	\item \textbf{pitch error rate} -- the relative error for the number of incorrectly detected pitches, $p_{\rm{incorrect}}$, excluding the extra detected notes:
\begin{displaymath}
\varepsilon_{\rm{pitch}} = \frac{p_{\rm{incorrect}}}{n_{\rm{original}}} \times 100 \%
\end{displaymath}
	\item \textbf{beat error rate} -- the relative error for the number of incorrectly detected beats, $b_{\rm{incorrect}}$, excluding the extra detected notes:
\begin{displaymath}
\varepsilon_{\rm{beat}} = \frac{b_{\rm{incorrect}}}{n_{\rm{original}}} \times 100 \%
\end{displaymath}
\end{itemize}

\section{Results}
\subsection{MuseScore generated melodies}

The number of notes of the original and automatically generated scores, and the number of incorrect pitches and beats excluding the extra detected notes, for each melody, are given in Table~\ref{table_errors_musescore}.

\begin{figure}[bt]
\centering
\vspace{-10pt}
\includegraphics[clip, width=3.5in]{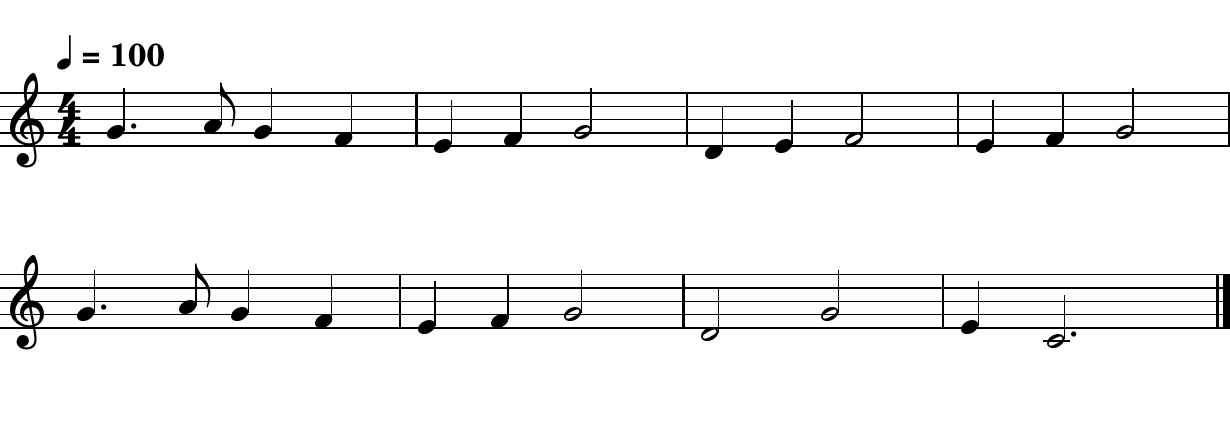}
\includegraphics[clip, width=3.5in]{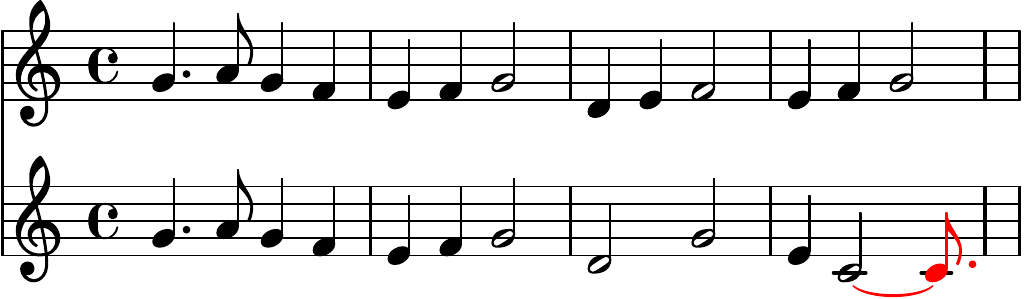}
\caption{``London Bridge'' original score (top) and generated score (bottom).}
\label{fig_london_bridge_musescore}
\vspace{-2pt}
\end{figure}

The chosen parameter values give nearly perfect results for the chosen melodies.
The incorrectly detected beats in the melodies mainly appear for the end note, as shown in Fig.~\ref{fig_london_bridge_musescore} for ``London Bridge''.
The errors in the generated score are marked with red.

From a musical standpoint, the last note marked with red in the generated score, which is tied with the previous note and has the same pitch, means that the duration of the previous note is increased, so the total beat duration of the tie is 2$\frac{3}{4}$.
In fact, the system detects the last note as one note with duration 2$\frac{3}{4}$, but the score generator renders it as a tie.

\begin{table}[bt]
\renewcommand{\arraystretch}{2.0}
\caption{Comparisons of the original and generated music scores for melodies generated using \emph{MuseScore}.}
\label{table_errors_musescore}
\centering
\begin{tabular}{|c|c|c|c|c|}
\hline
\bfseries \makecell{Melody name} & \bfseries \makecell{Number of\\notes\\ (original)} & \bfseries \makecell{Number of\\notes\\(detected)} & \bfseries \makecell{Incorrect\\pitches} & \bfseries \makecell{Incorrect\\beats}\\
\hline
\makecell{The Alphabet\\song} & 43 & 43 & 0 & 0\\
\hline
\makecell{Auld Lang\\Syne} & 58 & 58 & 0 & 1\\
\hline
\makecell{Cannon in D} & 46 & 46 & 0 & 0\\
\hline
\makecell{Happy Birthday} & 25 & 25 & 0 & 1\\
\hline
\makecell{Jingle Bells} & 49 & 49 & 0 & 1\\
\hline
\makecell{London Bridge} & 24 & 25 & 0 & 0\\
\hline
\makecell{Mary had a\\little lamb} & 25 & 25 & 0 & 1\\
\hline
\makecell{Ode to Joy} & 62 & 62 & 0 & 0\\
\hline
\makecell{Silent Night} & 47 & 47 & 0 & 0\\
\hline
\makecell {Twinkle\\Twinkle\\Little Star} & 42 & 42 & 0 & 0\\
\hline
\end{tabular}
\end{table}

\subsection{Digital piano recordings}

Table~\ref{table_errors_piano} shows the error rates for generated scores using Scorpiano and AnthemScore for the selected melodies.
The worst results using Scorpiano are obtained for the ``Silent Night'' melody.
The main reason for this is because the tempo of the melody is estimated to be 143 bpm, which is twice the actual of 70 bpm.
Because of this, every note in the generated score has nearly twice the actual beat duration.

\begin{table}[bt]
\renewcommand{\arraystretch}{2.0}
\caption{Error rates for the automatically generated scores using \emph{Scorpiano} and \emph{AnthemScore} for melodies played with the digital piano.}
\label{table_errors_piano}
\centering
\begin{tabular}{|c|c|c|c|c|}
\hline
\bfseries \makecell{Melody name} & \bfseries \makecell{Algorithm} & \bfseries \makecell{Note\\error rate} & \bfseries \makecell{Pitch\\error rate} & \bfseries \makecell{Beat\\error rate} \\
\hline
\multirow{2}{*}{\makecell{Auld Lang\\Syne}} & Scorpiano & 3.45 & 1.72 & 0.00 \\
\cline{2-5}
& AnthemScore & \bfseries 1.72 & \bfseries 0.00 & 0.00 \\
\hline
\multirow{2}{*}{\makecell{Canon in D}} & Scorpiano & \bfseries 4.35 & 0.00 & \bfseries 2.17 \\
\cline{2-5}
& AnthemScore & 8.70 & 0.00 & 56.52 \\
\hline
\multirow{2}{*}{\makecell{London Bridge}} & Scorpiano & 0.00 & 0.00 & 16.67 \\
\cline{2-5}
& AnthemScore & 0.00 & 0.00 & 16.67 \\
\hline
\multirow{2}{*}{\makecell{Ode to Joy}} & Scorpiano & \bfseries 0.00 & 3.23 & \bfseries 0.00 \\
\cline{2-5}
& AnthemScore & 17.74 & \bfseries 0.00 & 1.61 \\
\hline
\multirow{2}{*}{\makecell{Silent Night}} & Scorpiano & 70.21 & 0.00 & 100 \\
\cline{2-5}
& AnthemScore & \bfseries 8.51 & 0.00 & \bfseries 0.00 \\
\hline
\end{tabular}
\end{table}

\begin{figure}[bt]
\centering
\vspace{-10pt}
\includegraphics[clip, width=3.5in]{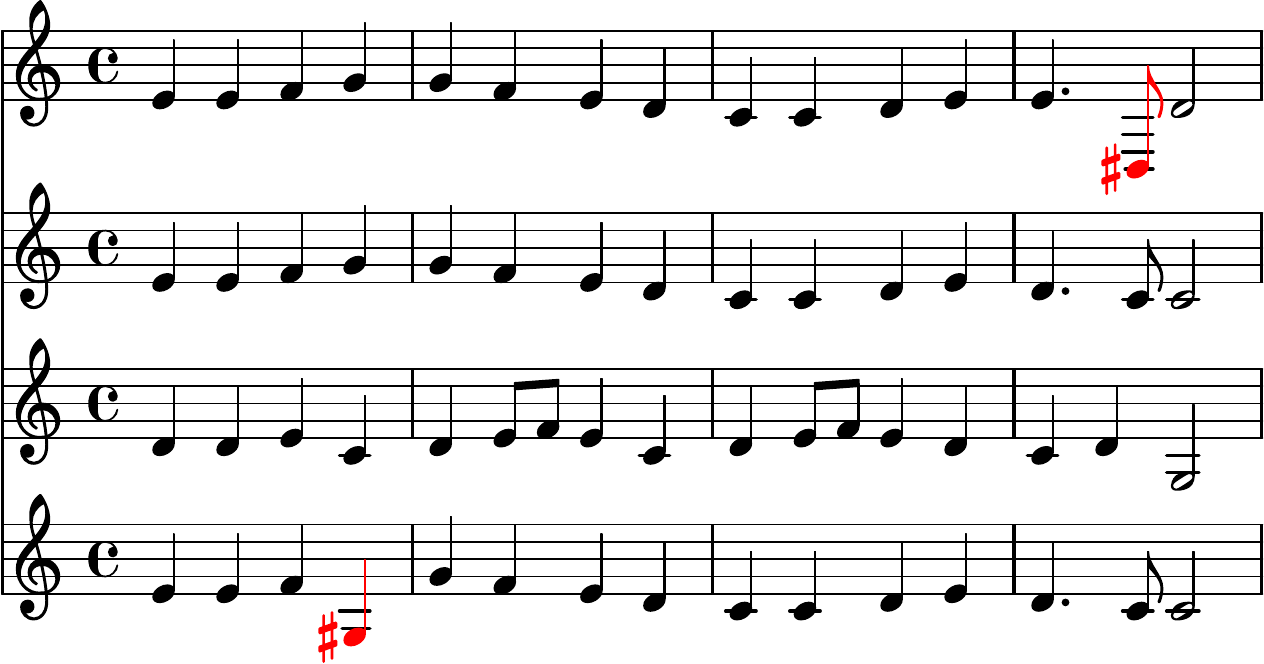}
\includegraphics[clip, width=3.5in]{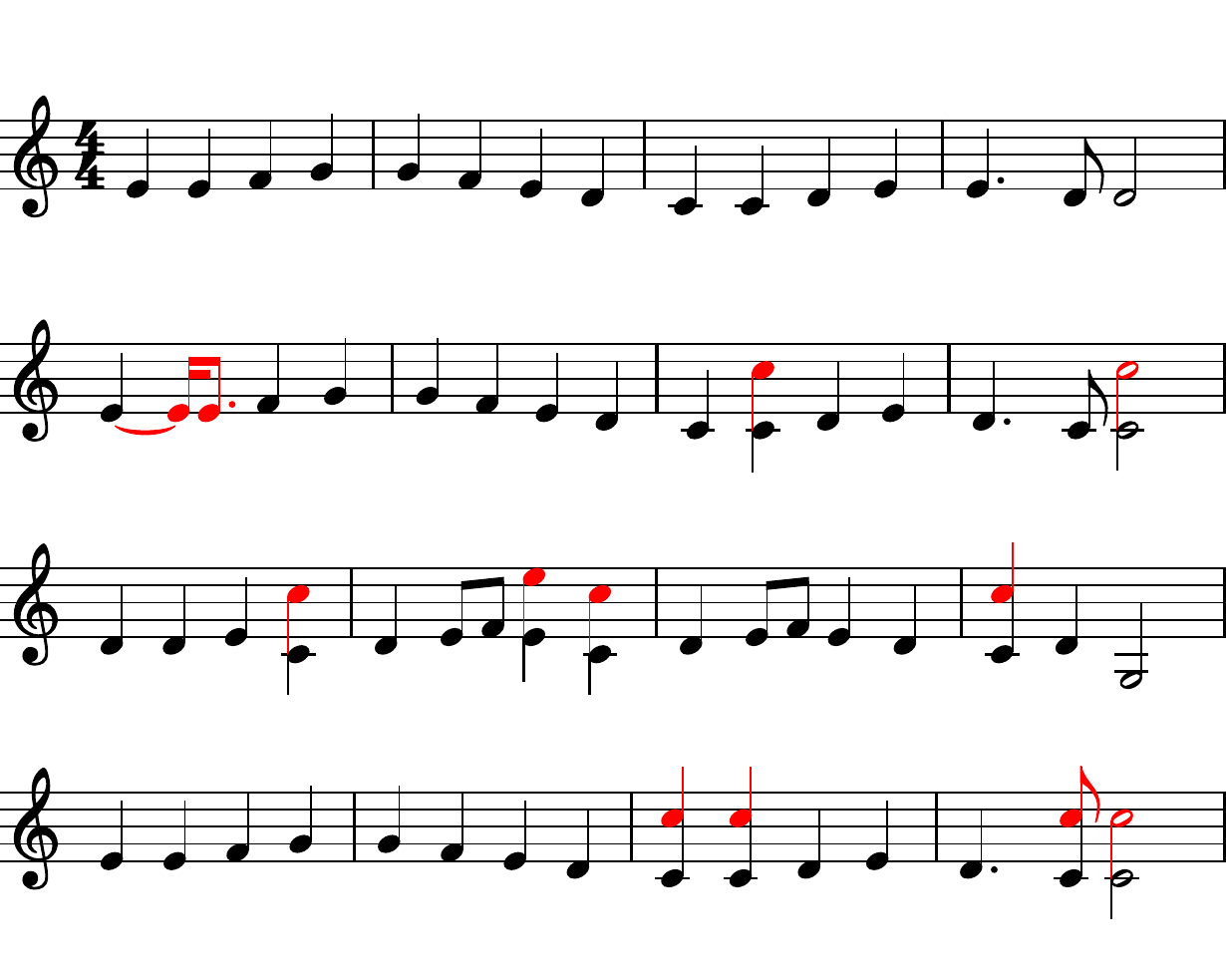}
\caption{Scorpiano (top) and AnthemScore (bottom) automatically generated scores for ``Ode to Joy'' played with a digital piano.}
\label{fig_ode_joy_piano}
\vspace{+2pt}
\end{figure}

The errors in the scores generated with AnthemScore mostly comprise extra notes that are detected as played simultaneously with the correct note, forming a chord.
The worst results are for ``Canon in D'' melody for which AnthemScore estimated the tempo incorrectly to be 113 bpm, versus the actual of 76 bpm.
Despite of this, AnthemScore gives really good results for the digital piano recordings, show casing the advantage of using neural networks for AMT.
Fig.~\ref{fig_ode_joy_piano} shows a comparison of the errors in transcription for ``Ode to Joy'' generated with Scorpiano and AnthemScore.
The notes colored red in the generated score with Scorpiano have incorrectly detected pitch.
Their fundamental frequency was estimated to be two times smaller than the actual one.
This type of error in pitch detection, when the estimated fundamental frequency and the actual one form a ratio equal to a power of 2 is known as \emph{octave error}\cite{OCTERR}.

\subsection{Tempo effect}

Table~\ref{table_errors_tempo} shows the average error rates for generated scores with Scorpiano, for the melodies ``Jingle Bells'' and ``Twinkle Twinkle Little Star'' played on a digital piano with three different tempos: slow, normal and fast.
From the results, it can be concluded that overall, the tempo has a small effect on the system performance, although there are cases, like with the ``Silent Night'' melody in the digital piano experiment, where the incorrect estimation of the tempo makes beat duration of the notes incorrect.
This problem also happened with AnthemScore in the digital piano experiment for ``Canon in D'' melody.

\begin{table}[bt]
\renewcommand{\arraystretch}{2.0}
\caption{Error rates for generated scores using \emph{Scorpiano} for melodies played on a digital piano with three different tempos.}
\label{table_errors_tempo}
\centering
\begin{tabular}{|c|c|c|c|}
\hline
\bfseries \makecell{Tempo} & \bfseries \makecell{Note\\error rate} & \bfseries \makecell{Pitch\\error rate} & \bfseries \makecell{Beat\\error rate} \\
\hline
\makecell{Slow} & 0.00 & 0.00 & 1.02 \\
\hline
\makecell{Normal} & 0.00 & 0.00 & 1.02 \\
\hline
\makecell{Fast} & 2.38 & 0.00 & 0.00 \\
\hline
\end{tabular}
\end{table}

\subsection{Summary}

Table \ref{table_errors_summary} summarises the average error rates for Scorpiano and AnthemScore for each experiment.
Both systems score equally good for the MuseScore and Tempo experiment.
AnthemScore shows slightly better results for the Digital piano experiment.

\begin{table}[b]
\renewcommand{\arraystretch}{2.0}
\caption{Summary results.}
\label{table_errors_summary}
\centering
\begin{tabular}{|c|c|c|c|c|}
\hline
\bfseries \makecell{Experiment} & \bfseries \makecell{Algorithm} & \bfseries \makecell{Note\\error rate} & \bfseries \makecell{Pitch\\error rate} & \bfseries \makecell{Beat\\error rate}\\
\hline
\multirow{2}{*}{\makecell{MuseScore}} & \bfseries Scorpiano & \bfseries 0.42 & \bfseries 0.00 & \bfseries 1.18 \\
\cline{2-5}
& AnthemScore & 0.99 & 0.00 & 1.49 \\
\hline
\multirow{2}{*}{\makecell{Digital piano}} & Scorpiano & 8.20 & 0.89 & 12.89 \\
\cline{2-5}
& \bfseries AnthemScore & \bfseries 5.94 & \bfseries 0.00 & \bfseries 8.35 \\
\hline
\multirow{2}{*}{\makecell{Tempo}} & \bfseries Scorpiano & \bfseries 0.79 & \bfseries 0.00 & \bfseries 1.02 \\
\cline{2-5}
& AnthemScore & 1.70 & 0.00 & 1.19 \\
\hline
\end{tabular}
\end{table}

\section{Conclusion}
The problem of AMT for monophonic music can be effectively addressed using digital processing techniques.
We propose a AMT system for monophonic piano music, that uses pure digital signal processing and has the advantage of being computationally inexpensive, fast, and does not need big training sets, whilst obtaining good results with low error rates, comparable to commercial neural network based systems.

In its current form, the system lacks detection of breaks and time signatures.
Sometimes mistakes can make the generated score look wrong, although most of the time the errors were obvious and could easily be corrected by human intervention.

The described system can be extended in the future.
Postprocessing the outputs of the modules can improve the performance of the system.
For example, the estimated fundamental frequency for each note can be compared with the neighbouring notes to avoid octave errors.
The system can also be modified to work with different musical instruments.

\bibliographystyle{IEEEtran}
\bibliography{IEEEabrv,ANT}

\begin{thebibliography}{10}
\providecommand{\url}[1]{#1}
\csname url@samestyle\endcsname
\providecommand{\newblock}{\relax}
\providecommand{\bibinfo}[2]{#2}
\providecommand{\BIBentrySTDinterwordspacing}{\spaceskip=0pt\relax}
\providecommand{\BIBentryALTinterwordstretchfactor}{4}
\providecommand{\BIBentryALTinterwordspacing}{\spaceskip=\fontdimen2\font plus
\BIBentryALTinterwordstretchfactor\fontdimen3\font minus
  \fontdimen4\font\relax}
\providecommand{\BIBforeignlanguage}[2]{{%
\expandafter\ifx\csname l@#1\endcsname\relax
\typeout{** WARNING: IEEEtran.bst: No hyphenation pattern has been}%
\typeout{** loaded for the language `#1'. Using the pattern for}%
\typeout{** the default language instead.}%
\else
\language=\csname l@#1\endcsname
\fi
#2}}
\providecommand{\BIBdecl}{\relax}
\BIBdecl

\bibitem{AMT:1}
E.~Benetos, S.~Dixon, D.~Giannoulis, H.~Kirchhoff, and A.~Klapuri, ``Automatic
  music transcription: Challenges and future directions,'' \emph{Journal of
  Intelligent Information Systems}, vol.~41, 12 2013.

\bibitem{AMT:2}
E.~Benetos, S.~Dixon, Z.~Duan, and S.~Ewert, ``Automatic music transcription:
  An overview,'' \emph{IEEE Signal Processing Magazine}, vol.~36, no.~1, pp.
  20--30, 2019.

\bibitem{AMTpoly:1}
G.~Costantini, M.~Todisco, and G.~Saggio, ``Automatic music transcription based
  on non-negative matrix factorization,'' 2010.

\bibitem{AMTpoly:2}
J.~Sleep, ``Automatic music transcription with convolutional neural networks
  using intuitive filter shapes,'' 10 2017.

\bibitem{PITCHm:1}
P.~S. Rao, S.~Khoushikh, S.~Ravishankar, R.~A. Ananthkrishnan, and
  K.~Balachandra, ``A comparative study of various pitch detection
  algorithms,'' in \emph{2020 5th International Conference on Computing,
  Communication and Security (ICCCS)}, 2020, pp. 1--6.

\bibitem{YIN:1}
A.~de~Cheveigné and H.~Kawahara, ``Yin, a fundamental frequency estimator for
  speech and music,'' \emph{Acoustical Society of America}, vol.~13, Apr. 2002.

\bibitem{PITCHf:1}
B.~Gerazov and Z.~Ivanovski, ``Building a basis for automatic melody extraction
  from macedonian rural folk music,'' 06 2010.

\bibitem{ONSET:2}
\BIBentryALTinterwordspacing
C.~M.~T. Rosão, ``Onset detection in music signals,'' Ph.D. dissertation,
  University of Lisbon, 2012. [Online]. Available:
  \url{http://hdl.handle.net/10071/5991}
\BIBentrySTDinterwordspacing

\bibitem{AMTML:1}
\BIBentryALTinterwordspacing
F.~Saputra, U.~G. Namyu, Vincent, D.~Suhartono, and A.~P. Gema, ``Automatic
  piano sheet music transcription with machine learning,'' \emph{Journal of
  Computer Science}, vol.~17, no.~3, pp. 178--187, Mar. 2021. [Online].
  Available: \url{https://thescipub.com/abstract/jcssp.2021.178.187}
\BIBentrySTDinterwordspacing

\bibitem{ONSET:1}
J.~P. Bello, L.~Daudet, S.~Abdallah, C.~Duxbury, M.~Davies, and M.~B. Sandler,
  ``A tutorial on onset detection in music signals,'' \emph{IEEE TRANSACTIONS
  ON SPEECH AND AUDIO PROCESSING}, vol.~13, pp. 1035--1047, Sep. 2005.

\bibitem{LIBROSA}
B.~McFee, C.~Raffel, D.~Liang, D.~P. Ellis, M.~McVicar, E.~Battenberg, and
  O.~Nieto, ``librosa: Audio and music signal analysis in python,'' in
  \emph{Proceedings of the 14th python in science conference}, vol.~8, 2015.

\bibitem{OCTERR}
\BIBentryALTinterwordspacing
B.~Kumaraswamy and P.~G. Poonacha, ``Octave error reduction in pitch detection
  algorithms using fourier series approximation method,'' \emph{IETE Technical
  Review}, vol.~36, no.~3, pp. 293--302, 2019. [Online]. Available:
  \url{https://doi.org/10.1080/02564602.2018.1465859}
\BIBentrySTDinterwordspacing

\bibitem{MllerAUDIO}
M.~Mller, \emph{Fundamentals of Music Processing: Audio, Analysis, Algorithms,
  Applications}, 1st~ed.\hskip 1em plus 0.5em minus 0.4em\relax Springer
  Publishing Company, Incorporated, 2015.

\bibitem{Gao2015PITCHDB}
Q.~Gao, ``Pitch detection based monophonic piano transcription,''
  \emph{Yankee}, vol.~7, no.~60, p.~C4, 2015.

\end{thebibliography}

\nocite{MllerAUDIO}
\nocite{Gao2015PITCHDB}

\end{document}